\documentclass[prd, twocolumn, nofootinbib, floatfix]{revtex4-1}

\usepackage{amsmath}
\usepackage{graphicx}
\usepackage{dcolumn}
\usepackage{bm}
\usepackage{epsfig}
\usepackage{amssymb,latexsym,mathrsfs}
\usepackage{graphicx}
\usepackage{color}
\usepackage{hyperref}

\hypersetup{
    colorlinks=true,
    linkcolor=red,
    citecolor=blue,
} 

\newcommand{\be}{\begin{equation}}
\newcommand{\ee}{\end{equation}}

\newcommand{\bs}{\begin{split}} 
\newcommand{\bea}{\begin{eqnarray}}
\newcommand{\eea}{\end{eqnarray}}

\newcommand{\neff}{\Delta N_{\rm eff}}
\newcommand{\lcdm}{$\Lambda$CDM}

\begin{document}

\title{Dark Energy Scaling from Dark Matter to Acceleration} 
\author{Jannis Bielefeld$^1$, Robert R.\ Caldwell$^1$, Eric V.\ Linder$^{2,3}$} 
\affiliation{$^1$ Department of Physics \& Astronomy, Dartmouth College, 
Hanover, NH 03755 USA\\ 
$^2$ Berkeley Center for Cosmological Physics \& Berkeley Lab, 
University of California, Berkeley, CA 94720, USA\\ 
$^3$ Institute for the Early Universe WCU, Ewha Womans University, 
Seoul 120-750, Korea}

\begin{abstract}
The dark sector of the Universe need not be completely separable into distinct 
dark matter and dark energy components. 
We consider a model of early dark energy in which the dark energy 
mimics a dark matter component in both evolution and perturbations at early 
times. Barotropic aether dark energy scales as a fixed fraction, possibly 
greater than one, of the dark matter density and has vanishing sound speed 
at early times before undergoing a transition. 
This gives signatures not only in cosmic expansion but in sound speed and 
inhomogeneities, and in number of effective neutrino species. 
Model parameters describe the timing, sharpness of the transition, and the 
relative abundance at early times. Upon comparison with current data, we find 
viable regimes in which the dark energy behaves like dark matter at early times: 
for transitions well before recombination the dark energy to dark matter 
fraction can equal or exceed unity, while for transitions near recombination 
the ratio can only be a few percent. After the transition, dark energy goes 
its separate way, ultimately driving cosmic acceleration and approaching a 
cosmological constant in this scenario. 
\end{abstract}

\date{\today} 

\maketitle

\section{Introduction} 

The present energy budget of the Universe contains two significant 
components beyond the Standard Model of particle physics -- dark matter 
and dark energy -- with little definitely known about their nature.  
Today, they act very differently, with dark matter clustering into 
galaxies and enhancing gravitational attraction while dark energy appears 
spread nearly homogeneously throughout space with a tension that causes the
acceleration of the cosmic expansion.  Here we consider whether they might 
have been more closely related in the past. 

Such a concept might be realized if dark matter and dark energy arose from the 
same, or related high energy physics processes. Indeed, such connections 
might arise from decaying moduli in string theory, e.g.\ see \cite{1401.4364} 
for such a case connecting dark matter and dark radiation. Modified gravity 
can also cause evolution of the couplings to and between different sectors, 
e.g.\ the early transition model of \cite{1312.3361}. In particular, 
the dark sector could have a different character at high redshift, and 
dark energy could have contributed dynamically at early times, perhaps with 
density at certain epochs comparable to that of dark matter. 
Several high energy physics origins for dark energy, such as Dirac-Born-Infeld scalar fields \cite{dbi1,dbi2,dbi3} or 
dilatons \cite{dilaton1,dilaton2}, predict such early dark energy, and in 
many cases it acts in a decelerating manner, possibly scaling as the dominant component 
of energy density, or simply like dark matter.  Moreover, 
such models often involve a non-relativistic sound speed of perturbations. 
Thus, such cold, early dark energy can act substantially like cold dark matter. 

For probing the early universe, measurements of the cosmic microwave background (CMB) 
offer the best evidence, and have already been used to place percent-level 
limits on dark energy at recombination \cite{planck16,hls}.  Here 
we investigate whether viable models exist in which the early dark energy 
density at some prerecombination epoch can be of order (or even greater) 
than the dark matter density, while possessing many of the same properties. 

Since the dark energy should today be accelerating and fairly smooth, 
this requires an evolution in its behavior in both equation of state and 
sound speed.  Moreover, so as not to disagree with formation of galaxies 
and clusters, the dark energy must quickly fade away from the early universe 
into the matter dominated era where structure grows.  Recently, 
\cite{13052209} investigated a model where such a transition occurred after 
recombination.  However, they kept the fluctuation sound speed in the dark energy to be the speed of light, 
reducing the effect of perturbations, and adopted a purely phenomenological 
model for the density evolution. 

Because the model here behaves like dark matter in both the expansion and 
perturbations at times before the transition, then if the transition occurred 
after recombination such additional energy density would just look like added 
dark matter and be faced with the usual CMB constraints on the dark matter 
density. Therefore we concentrate on the more interesting case for our model 
of a prerecombination transition and adopt for the dark energy the 
barotropic aether \cite{scher04,linscher}, a model with useful and 
interesting properties. 

In Sec.~\ref{sec:baro} we explore the physical effects of the barotropic 
aether as it evolves from dark matter-like behavior at early times to cosmic 
acceleration at late times.  We confront the model with recent CMB data 
in Sec.~\ref{sec:cmb}, and discuss what this may teach us about the dark sector 
in Sec.~\ref{sec:concl}.

\section{From Dark Matter to Acceleration} \label{sec:baro} 

The general models we are interested in exploring here are ones where 
the dark energy resembles dark matter and can be a significant fraction of 
the dark matter density at early times, but accelerates the expansion 
and has $w\approx-1$ to accord with observations at late times.  In order 
for this to work, there must be a period where the dark energy density 
rapidly declines relative to the matter -- what \cite{13052209} called 
freeze out behavior -- in order to satisfy both CMB constraints on the 
dark matter density and satisfy the growth of structure constraints during the 
matter dominated era.  Essentially this means that the dark energy must 
gain a positive equation of state, such as during kination, a period of 
kinetic dominated dynamics where $w=+1$.  This will constrain the physical 
classes viable to produce this scenario.

\subsection{Models} 

We approach this problem through the barotropic aether 
model \cite{scher04,linscher}, which is well suited to these behaviors, 
rather than through attempts to actually unify dark energy and dark matter. 
Such unified models often 
have issues with incompleteness, e.g. how perturbations behave, fine tuning or difficulties matching observations. 

For example, one approach is to tailor a scalar field potential to give exactly the 
density evolution behavior desired.  Inverse power law potentials, for 
example, have attractor solutions where the dark energy density scales 
as some power of the expansion factor \cite{ratrap}.  The dark energy 
equation of state is $w=(nw_b-2)/(n+2)$ for power law index $-n$ and 
background equation of state $w_b$, so by taking $n=6$ one can arrange 
dark energy to track the matter density during radiation domination.  
However to institute the kination phase, the potential has to steepen 
drastically, {\`a} la slinky model \cite{Barenboim:2005np}, and then become shallow 
to allow for late time acceleration.  This seems quite fine tuned. 

A second approach is to allow explicit interactions between dark matter 
and dark energy.  Such a scenario can lead to a constant ratio between 
their densities, or at least a long period where they are comparable, 
but generally within the matter dominated rather than radiation dominated 
era.  An interaction term 
\be 
\Gamma=3wH\,\left[\frac{1}{\rho_{de}}+\frac{1}{\rho_{dm}}\right]^{-1} \ , 
\ee 
entering with opposite signs in the $\dot\rho_{de}$ and $\dot\rho_{dm}$ 
evolution equations, where $w$ is the bare equation of state of the dark 
energy and $H=\dot a/a$ is the Hubble parameter, 
will give $w_{de}=w_{dm}$.  But endowing dark matter with even a small amount of pressure tends to 
cause disagreements with observations, certainly during the matter dominated 
era where it causes a large integrated Sachs-Wolfe signal in the CMB.  
Furthermore, there is no clear mechanism to later obtain kination, and then 
acceleration, which would require $\Gamma$ to change sign.  This also 
introduces two arbitrary functions: the scalar field potential and the 
interaction. 

Another class of models that are related to the barotropic aether is based on phenomenological properties of the dark fluid. Interesting examples are the generalized Chaplygin gas \cite{Kamenshchik:2000iv} and the condensate cosmology \cite{Bassett:2002fe}. These are all motivated by unifying dark energy and dark matter to one dark fluid in the early universe. Physically, this seems 
problematic as these two dark species must simultaneously exist today, 
and have been invoked to explain very different and apparently incompatible phenomena. One provides extra gravitational attraction on small length scales, and the other seems to cause gravitational repulsion on large length scales. One appears to aggregate and clump, and the other appears to be very smooth. On the other hand, it seems reasonable to explore connections between the two species, such as interactions or a possible related origin in a dark sector.

One could also attempt to carry out the desired dynamical evolution  by 
changing the kinetic structure of the theory, i.e.\ using a k-essence 
model \cite{kess1,kess2}.  This also avoids the necessity for a potential 
(e.g.\ \cite{scher04,deputter}) and 
adds richness to the perturbation evolution by determining a time varying 
sound speed. However the structure of the kinetic function would need 
to be essentially as complicated as the scalar field potential in the 
quintessence approach.  The barotropic aether we next consider is closely related to 
k-essence but with a simpler structure and with desirable properties that 
ameliorate many of the issues.

\subsection{Barotropic Aether} 

The barotropic aether model has the advantages 
that it has some physical foundation -- 
the pressure is an explicit function of the energy density -- it can be 
viewed as a purely kinetic k-essence model or a quintessence model, and 
its phase space evolution is nontrivial, allowing a rapid freezeout of 
the aether component and approach to current acceleration. 

Barotropic models have the same number of degrees of freedom as quintessence 
models, but also allow the sound speed to differ from the speed of light. 
In contrast, a quintessence scalar field that scales like matter with 
vanishing pressure, $w=0$, does not cluster like dark matter since it 
still possesses a relativistic sound speed,
equal to the speed of light. For barotropic 
models, the sound speed is determined by the equation of state, or vice 
versa, and barotropic aether models can have both $w=0$ and $c_s=0$.
The dynamics of barotropic models is given by 
\be 
	\frac{d \, w}{d \ln a}=-3(1+w)(c_s^2-w) \ , 
\ee 
so we can write $w(c_s)$ or $c_s(w)$, i.e.\ we only have to specify 
one function. Note that for these dynamics we have $c_s^2 = \delta p / \delta \rho$. There is also an attractor solution to $w=-1$, exactly what 
we need for late time acceleration.  As pointed out by \cite{linscher}, 
the transition to $w=-1$ is quite rapid, usually within an e-fold, 
validating the observation of a present value close to $-1$. 

If we model the early time behavior such that $c_s\to0$ or $w\to0$ (one 
enforces the other), then we also have a partial solution to the coincidence 
problem, in that there can be a period with a constant ratio of dark energy to dark 
matter densities.  These characteristics -- early time scaling, rapid 
transition, late time $w=-1$ -- motivate consideration of the  barotropic 
aether early dark energy as an effective transition model. 

Since the late time behavior of the aether models converges on the de Sitter 
state with $w=-1$, we can consider the barotropic energy density as the 
sum of a constant and the deviations from the asymptotic density, 
$\rho_{de}=\rho_\infty+\rho_{ae}$.  Indeed 
\cite{linscher} showed that one can always split a barotropic model into 
a constant density piece and an ``aether'' piece that is itself barotropic 
and has positive equation of state $0\le w_{ae}\le1$.  
We therefore choose $w_{ae}$ to go from 0 in the past to 1 at 
later times, using the e-fold form 
\be 
w_{ae}(a)=\frac{1}{1+(a_t/a)^{1/\tau}} \ , \label{eq:wae} 
\ee 
where $a_t$ is the transition scale factor and $\tau$ is the rapidity of 
the transition in e-folds of expansion factor.  This then determines all 
the dynamics, with the total dark energy equation of state given by 
\bea 
w(a)&=&w_{ae}\,\frac{\rho_{ae}}{\rho_{de}}-\frac{\rho_\infty}{\rho_{de}}\\ 
&=&-1+(1+w_{ae})\,\frac{\rho_{ae}}{\rho_{de}} \ , 
\label{eq:wde}
\eea 
going from 0 in the distant past, to 1 just after the transition, to $-1$ 
at late times, 
and the sound speed given by 
\be 
c_s^2= \frac{1}{1+(a_t/a)^{1/\tau}}\,\left[1-\frac{1}{3\tau}\frac{1}{1+2(a/a_t)^{1/\tau}}\right] \ , 
\ee 
going from 0 at early times to 1 at late times.  

The form for $w_{ae}$ in Eq.~(\ref{eq:wae}) allows analytic calculation of 
the energy density 
\be 
\rho_{ae}(a)=\rho_{ae,0}\,a^{-3} \left(1+a_t^{-1/\tau}\right)^{3\tau} 
\left[1+\left(\frac{a}{a_t}\right)^{1/\tau}\right]^{-3\tau}  \label{eq:rhoae} 
\ee 
where the scale factor $a=1$ at the present day.
At early times this evolves as $a^{-3}$, leading to a constant density 
ratio $R=[\rho_{de}/\rho_m](a\ll a_t)$ relative to matter (independent of whether the expansion is 
dominated by radiation or matter), and then at times later than the transition 
it dies off quickly as $a^{-6}$, as for a free field.  In this 
sense, we regard $a_t$ as the transition to freeze out. 

\begin{figure}[htbp!]
\includegraphics[width=\columnwidth]{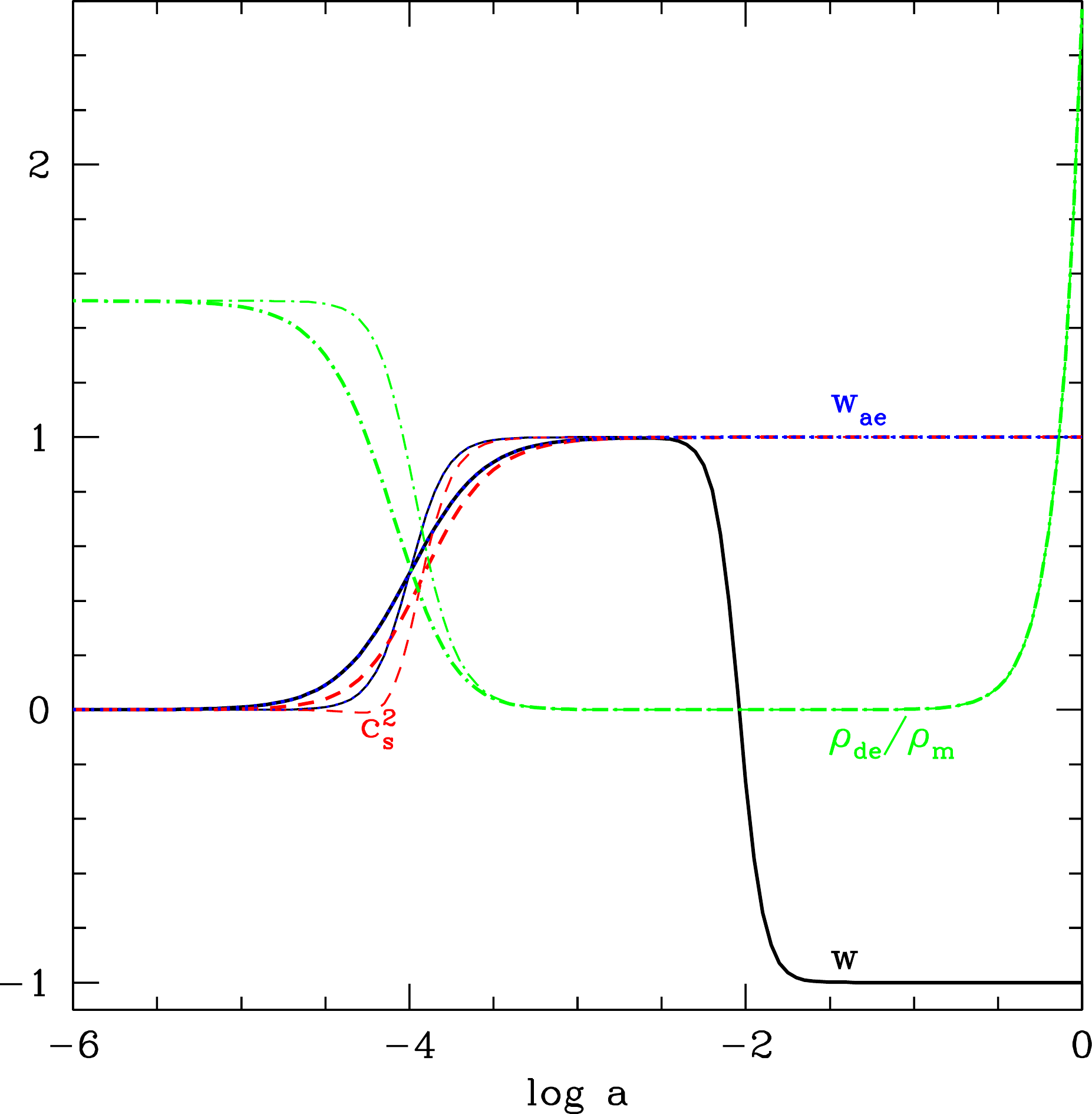} 
\caption{For the barotropic aether we plot the ratio of the total dark 
energy density to the matter density $\rho_{de}/\rho_m$, the total dark 
energy equation of state parameter $w$, the dark energy sound speed squared 
$c_s^2$, and the aether equation of state parameter $w_{ae}$.  Thick 
curves take $\tau=0.5$, thin curves $\tau=0.25$, with both having 
$a_t=10^{-4}$ and $R=\rho_{de}/\rho_m(a\ll a_t)=1.5$. 
} 
\label{fig:barow} 
\end{figure}

Figure~\ref{fig:barow} illustrates these behaviors.  We have three 
parameters: the time of transition $a_t$, width of transition $\tau$, 
and asymptotic early time ratio $R$ of dark energy to dark matter density.  
The last quantity also determines the present ratio of the aether piece to 
the constant density in the dark energy.

We see several interesting properties.  Due to the rapidity of the 
transition, if the transition takes place sufficiently before recombination 
then the dark energy does not affect the dark matter component at 
recombination, and the CMB power spectra should show little effect.  (Note 
that \cite{13052209} considered transitions after recombination to avoid 
disturbing the CMB power spectra.)  Nevertheless for the earlier history 
of the universe there can be a tie between the components of the dark 
sector, possibly alleviating the coincidence problem.  
While at late times, $w$ is extremely close to $-1$ and 
the $\Lambda$CDM model is an excellent approximation.  

However, signatures exist around the transition.  The sound speed begins 
to deviate from 0, as $w_{ae}$ deviates from 0, and so the dark energy 
acts differently from dark matter, changing the evolution of matter 
perturbations and the matter density power spectrum.  Indeed, the dark 
energy itself can start to cluster, but this is usually a smaller effect 
than its influence (through the gravitational potential) on the matter 
(see, e.g., \cite{rdpsound}).  For a rapid transition, with $\tau<1/3$, 
there is an epoch with $c_s^2<0$, which leads to instabilities in 
the dark energy clustering; therefore we only consider $\tau>1/3$. 
For slower transitions the distinction between the dark energy and
dark matter starts sooner and the dark energy density declines to lower values more quickly, 
leading to smaller signatures. For $\tau\gg1$, the model approaches 
$\Lambda$CDM. 

\begin{figure}[htbp!]
	\includegraphics[width=\columnwidth]{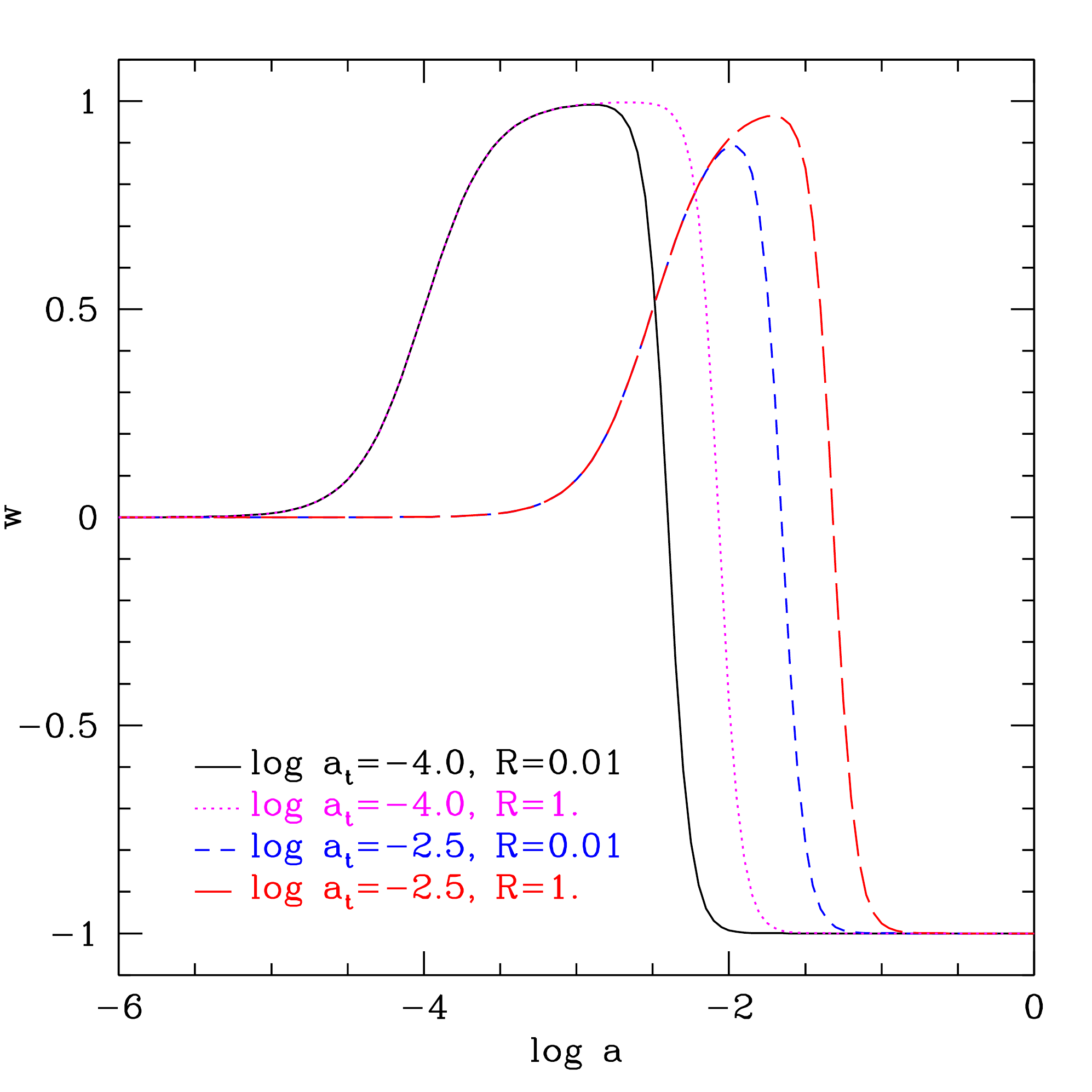} 
	\caption{The total dark energy equation of state $w(a)$ is plotted 
for several values of the transition scale factor $a_t$ and early dark 
energy-dark matter density ratio $R$ (for fixed $\tau=0.5$). The location 
of the transition away from dark matter behavior $w=0$ is determined by 
$a_t$ and the length of aether dominated behavior ($w=1$) by $R$, but at 
late times all curves go to the dark energy attractor with $w=-1$.} 
	\label{fig:barowfall} 
\end{figure}

It is instructive to analyze the dependence of the equation of state (EOS) of the aether, Eq.~(\ref{eq:wde}), on the model parameters. The transition scale factor $a_t$ determines the time when the EOS deviates from dark matter 
behavior $w=0$ and rises up to aether domination $w \to 1$. 
The subsequent rapid redshifting and drop to the dark energy attractor 
behavior $w\to -1$ gets delayed for large values of $R$. Nevertheless, 
as we later see, $R$ will generally be constrained to be at most a few 
percent for late time transitions, so the dark energy reaches its attractor 
behavior $w\to -1$ at redshifts well before the present. 
Figure~\ref{fig:barowfall} illustrates these dependences.

Moreover, the additional energy density of the early dark energy changes 
the expansion rate of the universe (assuming we do not reduce the dark 
matter density to compensate).  Around the time of the transition, the 
total dark energy equation of state $w$ rises through the radiation value 
of $1/3$ while the dark energy density is still appreciable.  This can 
be written in terms of a time varying, additional number of effective 
neutrino species: 
\be 
\neff= \left[\frac{7}{8}\left(\frac{4}{11}\right)^{4/3}\right]^{-1} 
\frac{\rho_{de}(a)}{\rho_\gamma(a)} \ , 
\ee 
where $\rho_\gamma$ is the photon energy density and the numerical factors 
arise from converting to effective neutrino species. 

\begin{figure}[htbp!]
\includegraphics[width=\columnwidth]{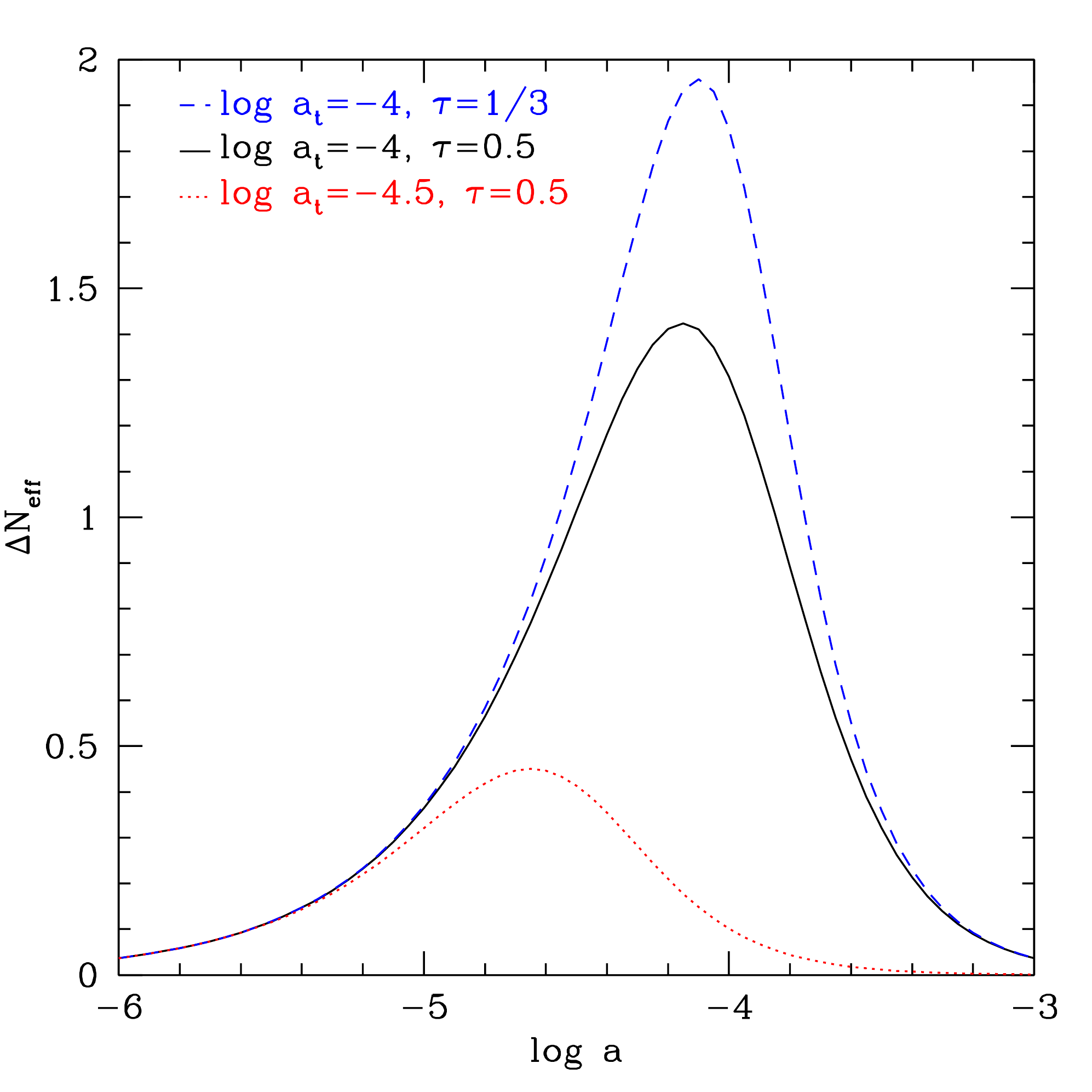} 
\caption{The effective number of extra neutrino species equivalent to 
the early dark energy density is plotted vs scale factor for different cases 
of the transition scale factor $a_t$ and width $\tau$, with fixed $R=1.5$. 
} 
\label{fig:neff} 
\end{figure}

Figure~\ref{fig:neff} shows the induced $\neff$ for various cases 
of transition time and width. The more 
rapid the transition (smaller $\tau$), the longer the dark energy density 
has been preserved and so the larger $\neff$. Holding the ratio $R$ fixed 
but moving the transition earlier has the effect of lowering $\neff$ since the radiation 
density was higher at those early times; conversely a later transition would 
enhance the bump in $\neff$. Finally, the amplitude of the bump scales 
linearly with $R$. CMB data should be sensitive to the value of $\neff$ near 
recombination. In particular, note that 
\be 
\neff\approx \left(\frac{a}{10^{-3}}\right) \, 
\left(\frac{\rho_{de}(a)/\rho_m(a)}{0.04}\right) \ . \label{eq:neffa} 
\ee

The perturbation equations for the barotropic aether are all standard and follow from the Einstein equations by including the property of barotropy that $\delta P / \delta \rho = c_s^2$. With this, in the synchronous gauge using the definitions from \cite{Ma:1995ey} we get
\begin{equation*}
	\delta' = -(1+w)\left(\theta + \frac{h'}{2}\right)-3\frac{a'}{a}\left(c_s^2 - w\right)\delta
\end{equation*}
together with
\begin{equation*}
	\theta' = -\frac{a'}{a}(1-3w)\theta - \frac{w'}{1+w}\theta + \frac{c_s^2}{1+w}k^2 \delta
\end{equation*}
where derivatives are taken with respect to conformal time. The matter density perturbation is defined by $\delta = \delta \rho / \rho$ and $\theta$ is the divergence of the fluid velocity $\theta = \nabla^j v_j$, while $h$ is the 
trace of the metric perturbation. 

\begin{figure}[h!]
\includegraphics[width=\columnwidth]{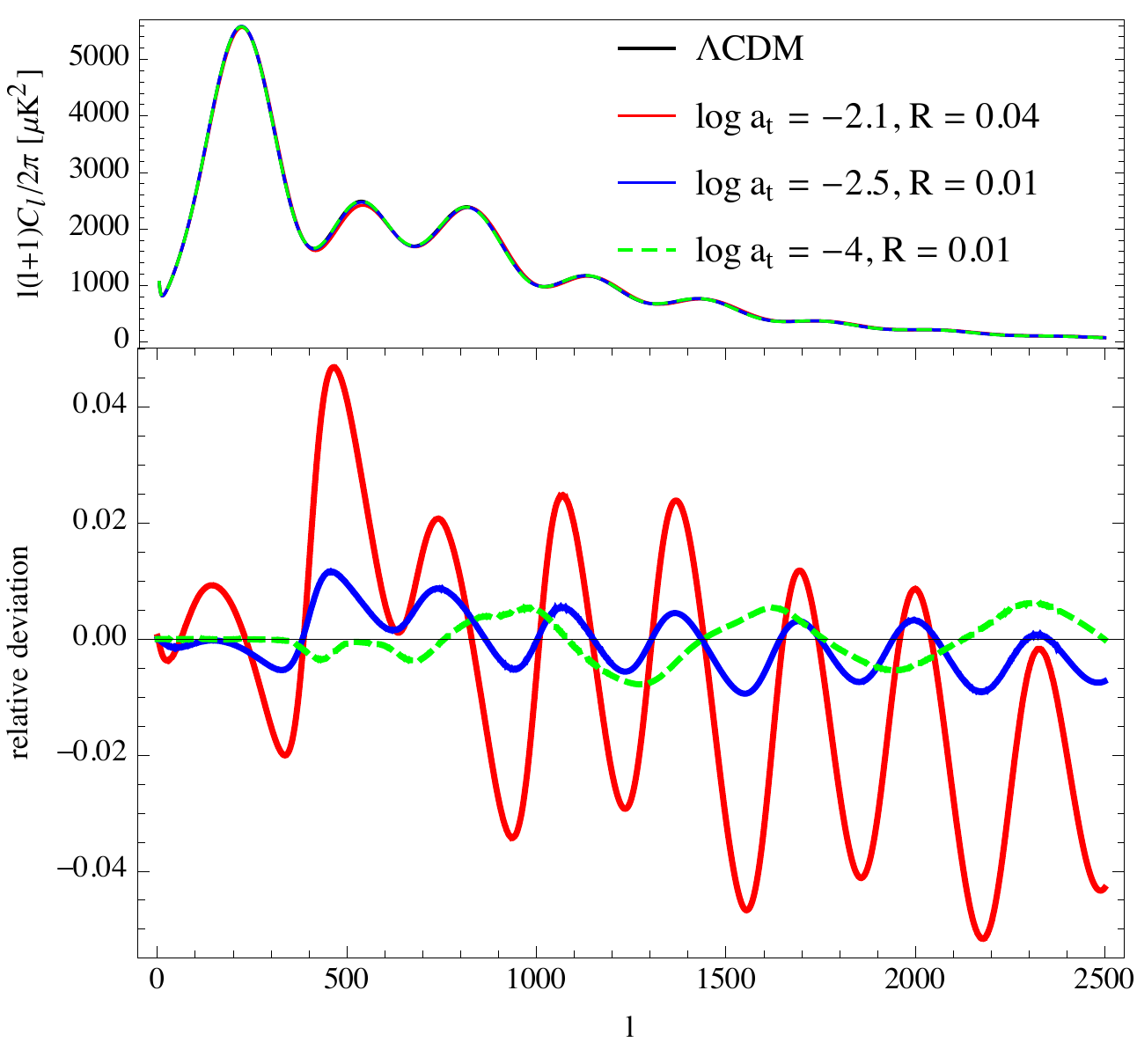} 
\caption{The CMB temperature power spectrum for $\Lambda$CDM and several 
barotropic aether models are shown (top panel) along with the relative 
deviations from $\Lambda$CDM (bottom panel). Post-recombination transitions 
can be readily distinguished while subpercent level pre-recombination 
models are consistent with data. For this plot we fix the other cosmological 
parameters and set $\tau=1$.} 
\label{fig:relerr} 
\end{figure}

The effect of the barotropic aether on the CMB power spectrum is displayed in Fig.~\ref{fig:relerr}, for various choices of parameters that we will later 
see give deviations from $\Lambda$CDM at between 68--95\% confidence level 
for current data. Deviations in the power spectrum at the 1\% level, too 
small to be seen by eye, can still be distinguished by data. Note that a 
post-recombination transition affects all acoustic peaks due to the 
geometric shift and can easily be detected, while pre-recombination 
transitions have more influence on the higher multipole damping tail.

\section{Constraints from Data} \label{sec:cmb} 

The dark energy component affects the CMB fluctuations through changing 
the expansion rate (including the time of ``matter''-radiation equality 
and effective relativistic degrees of freedom) and perturbation evolution. 
The dominant effect tends to be from the expansion rate and so depends mostly 
on the dark energy density contribution. From Eq.~(\ref{eq:rhoae}) the energy density
has a nonlinear dependence on the transition location $a_t$ and width $\tau$, 
while scaling linearly with $R$. We can relate the dark energy to matter 
density ratio at CMB last scattering to its early time asymptote by 
\be 
\frac{\rho_{de}(a_{\rm lss})}{\rho_m(a_{\rm lss})}  = R\, \left[1+\left(\frac{a_{\rm lss}}{a_t}\right)^{1/\tau}\right]^{-3\tau} . \label{eq:rholss} 
\ee 

\begin{figure}[htbp!]
\includegraphics[width=\columnwidth]{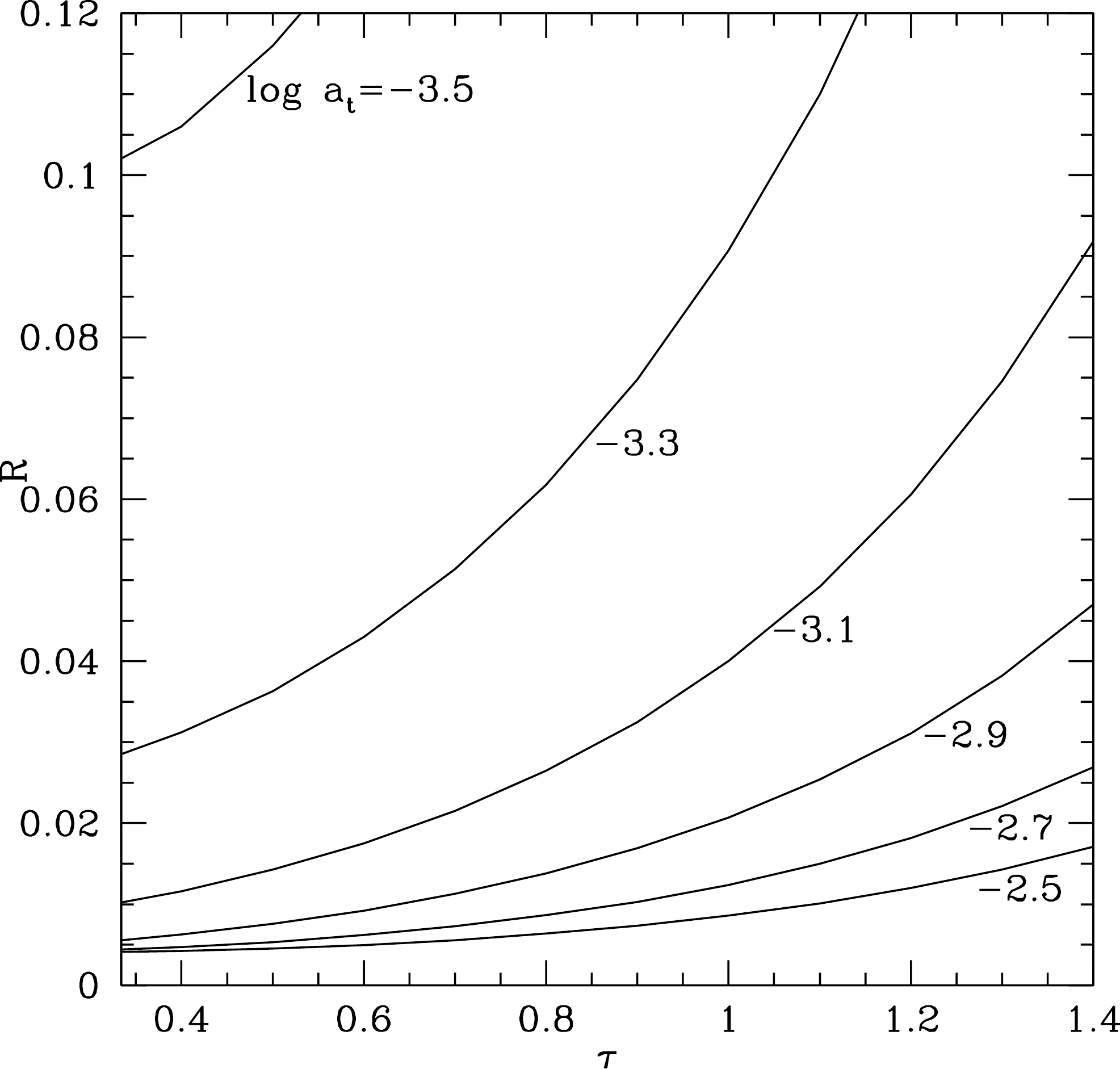} 
\caption{Assuming a bound on the dark energy to dark matter density at 
recombination imposes constraints on the dark energy parameters. Here for 
an upper bound of 0.4\%, only the parameter space below the respective 
$\log a_t$ curves is allowed. 
} 
\label{fig:rlss} 
\end{figure}

As a guide, if we want to keep the dark energy contribution at last scattering to below 
some number, say 0.4\% as a rough limit from Planck (\cite{planck16}, for a 
different, specific early dark energy model), then this defines an allowed 
region in the $R$-$a_t$-$\tau$ space. Figure~\ref{fig:rlss} illustrates this 
region in the $R$-$\tau$ plane for various slices of $a_t$.

The area below 
each curve is allowed, and we see that as the transition moves to much earlier 
times before recombination ($\log a_{\rm lss}=-3.04$) then much larger values of 
the asymptotic dark energy to dark matter density ratio $R$ are permitted. 
Indeed for $a_t\ll a_{\rm lss}$ the ratio $R$ grows as $a_t^{-3}$ so for 
$\log a_t=-3.9$ ($-4.3$) we can have early dark energy dominating dark matter 
by $R$ up to 1.5 (24), for any valid value of $\tau>1/3$, and still expect 
to have good agreement with CMB measurements. 

For robust constraints we modify CAMB inside CosmoMC \cite{Lewis:2002ah} to include the barotropic dark energy component. To constrain cosmology, including the new parameters of this theory $\{R, \tau, a_t\}$, 
we use CMB data from the Planck satellite \cite{Ade:2013kta} including the 
lensing potential. We complement the high-multipole tail with ACT and SPT 
data \cite{Keisler:2011aw,Reichardt:2011fv,Sievers:2013ica} extending 
up to multipole $l \sim 3000$. To help break other degeneracies we use the Hubble constant constraints from 
HST data \cite{Riess:2011yx}. Large scale structure data is included from 
the WiggleZ \cite{Blake:2011en} survey with GiggleZ corrections 
\cite{Parkinson:2012vd}, along with BAO data from the Sloan Digital Sky 
Survey (SDSS) DR9, DR7 \cite{Ahn:2012fh, Abazajian:2008wr} and the 
Anglo-Australian Observatory (AAO) 6DF survey \cite{Jones:2009yz}. 
Supernova data is taken from the Union 2.1 sample \cite{Suzuki:2011hu}. 

Our Markov Chain Monte Carlo (MCMC) analysis has 40 parameters, including 
the 6 standard cosmology parameters (a flat universe is assumed), 3  dark energy parameters, 
and the rest deal with instrumental and foreground effects. We let five chains run independently from each other to check convergence using the Gelman and Rubin (variance of chain means)/(mean of chain variances) $R_{GR}$ statistic \cite{Gelman:1992zz}. At the end of the MCMC run the worst-performing parameter with respect to this statistic, $a_t$, reached $R_{GR}=0.00115$. This demonstrates excellent chain convergence.

Since for early $a_t$ the constraints loosen drastically, as implied by Fig.~\ref{fig:rlss}, in our MCMC 
fitting of the model to data we will only consider the 
range $\log a_t\in [-3.3,-2.0]$. We do not extend the range to later 
times since the growth of cosmic structure rather than the CMB will impose 
the main constraints there, and the trend is already clear as discussed later 
in this section; furthermore similar late-time constraints were shown by 
\cite{13052209} for 
another transition model. Given that large $\tau$ moves the model closer 
and closer to $\Lambda$CDM, we only consider the range $\tau\in[0.33,1.2]$. 
Finally, because we have no a priori expected value for $R$, we use a 
logarithmic range of $\log R\in[-3.0,-1.3]$. We only expect larger $R$ to be 
allowed in the presence of large $\tau$ or early $a_t$, where constraints are 
weak. 
We have also tested wider prior bounds on $\log a_t$ and confirmed that 
the interesting behavior happens in the range used above; for early $a_t$ 
there is no constraining power since even large $R$ (initial dark energy 
to dark matter ratio) fades by the time of recombination. 

\begin{figure*}[!thbp]
	\includegraphics[width=\textwidth]{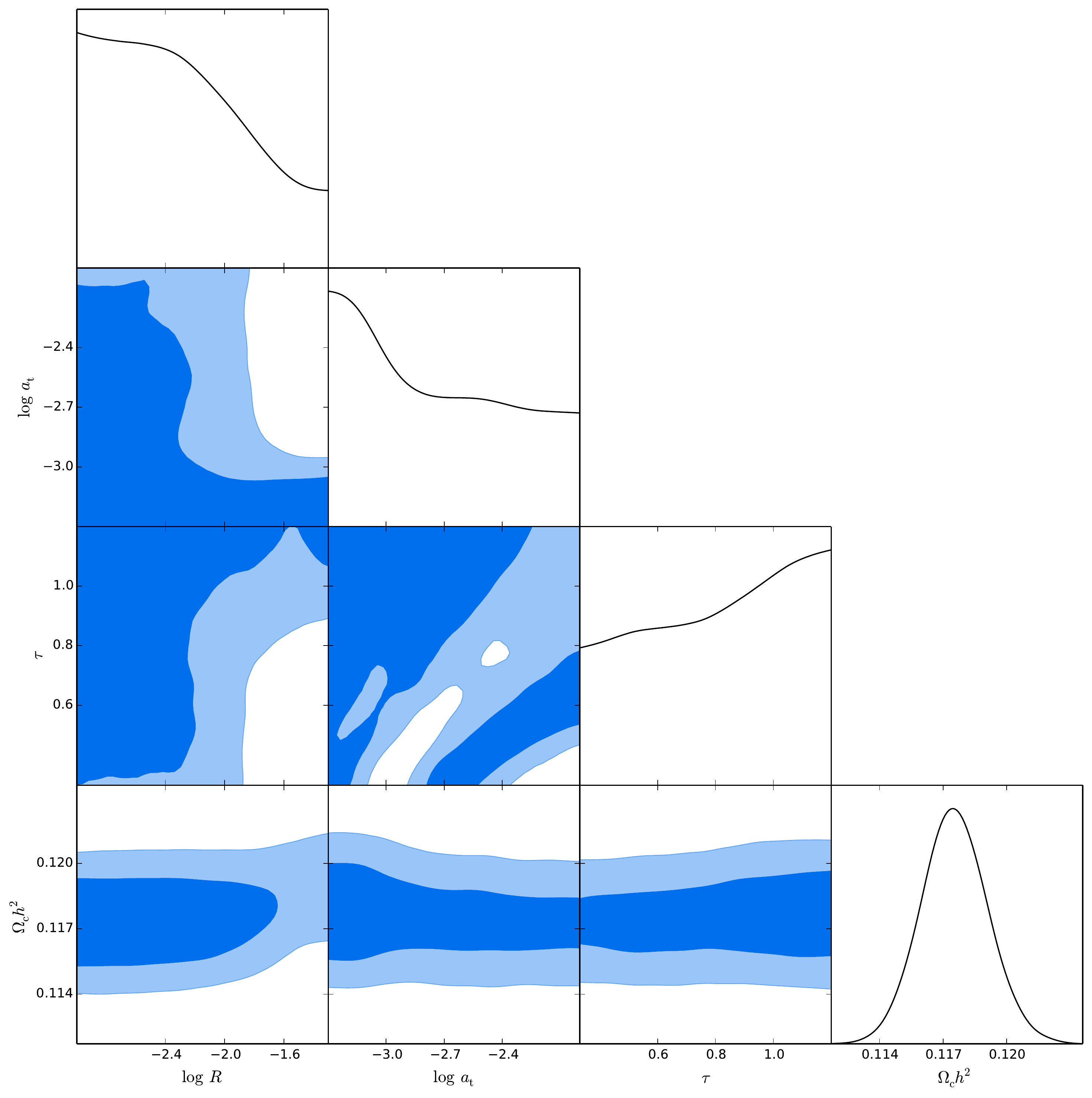} 
	\caption{2D joint confidence contours at 68\% CL (dark) and 
95\% CL (light) are shown for the dark energy and dark matter parameters, 
marginalized over all other parameters. The fully marginalized 1D PDFs 
are shown on the diagonal. 
	} 
	\label{fig:plot_tri} 
\end{figure*}

Figure~\ref{fig:plot_tri} shows the dark energy and dark matter joint 
constraints. 
The physics behind the dark energy influence can be clearly traced. 
Low values of $R$, early values of $a_t$, and large values of $\tau$ are 
all mildly preferred. These each push the model in a direction consistent with 
$\Lambda$CDM. The best fit parameters for this dark energy model give a 
likelihood with $\Delta\chi^2=6$ below \lcdm, with 3 more 
parameters.

First consider transitions earlier than the recombination epoch. The data 
has little constraining power because even for high values of $R$ a wide 
range of $\tau$ is allowed that gives sufficient time for the dark energy  
to fade away by recombination. At these early times, the dark 
energy is mimicking dark matter but less perfectly as the transition time 
approaches. As its sound speed and equation of state climbs above zero, 
this starts to cause decay of early dark matter gravitational potentials; 
to compensate for this (in the brief period before the dark energy fades 
to insignificance) the dark matter density needs to be slightly higher. 

For later transitions, the fit of  this dark energy model to the data holds 
only for progressively smaller values of $R$. For modest values of $R$ there 
is a complicated interplay between the dark energy density at last scattering 
and its behavior ($w$ and $c_s$). If the transition away from dark matter 
behavior ($w=0$, $c_s=0$) starts to happen before recombination, and there is 
sufficient dark energy density for this deviation to have physical effect, 
then the data disfavor this behavior. At a given $\log a_t$ after recombination, 
large $\tau$ causes the dark energy behavior to deviate before recombination, 
but the dark energy density fades more quickly (see Fig.~\ref{fig:barow}), 
and the model may be viable. Conversely, small $\tau$ preserves the matter-like dark energy 
behavior at recombination, and again the model can 
survive. However, for moderate $\tau$ the rate of fall of the dark energy 
density and the rise of the deviation in $w$ and $c_s$ can balance 
sufficiently to have an appreciable effect on the CMB (decaying  
gravitational potentials), which is disfavored 
by the data. Since the deviation occurs roughly at $\log a\approx\log a_t+\tau$, 
this region (for sufficiently large $R$) is disfavored, leading to the 
``blank stripe'' seen in the joint confidence contour of $\log a_t$--$\tau$. 

\begin{figure}[!thbp]
	\includegraphics[width=\columnwidth]{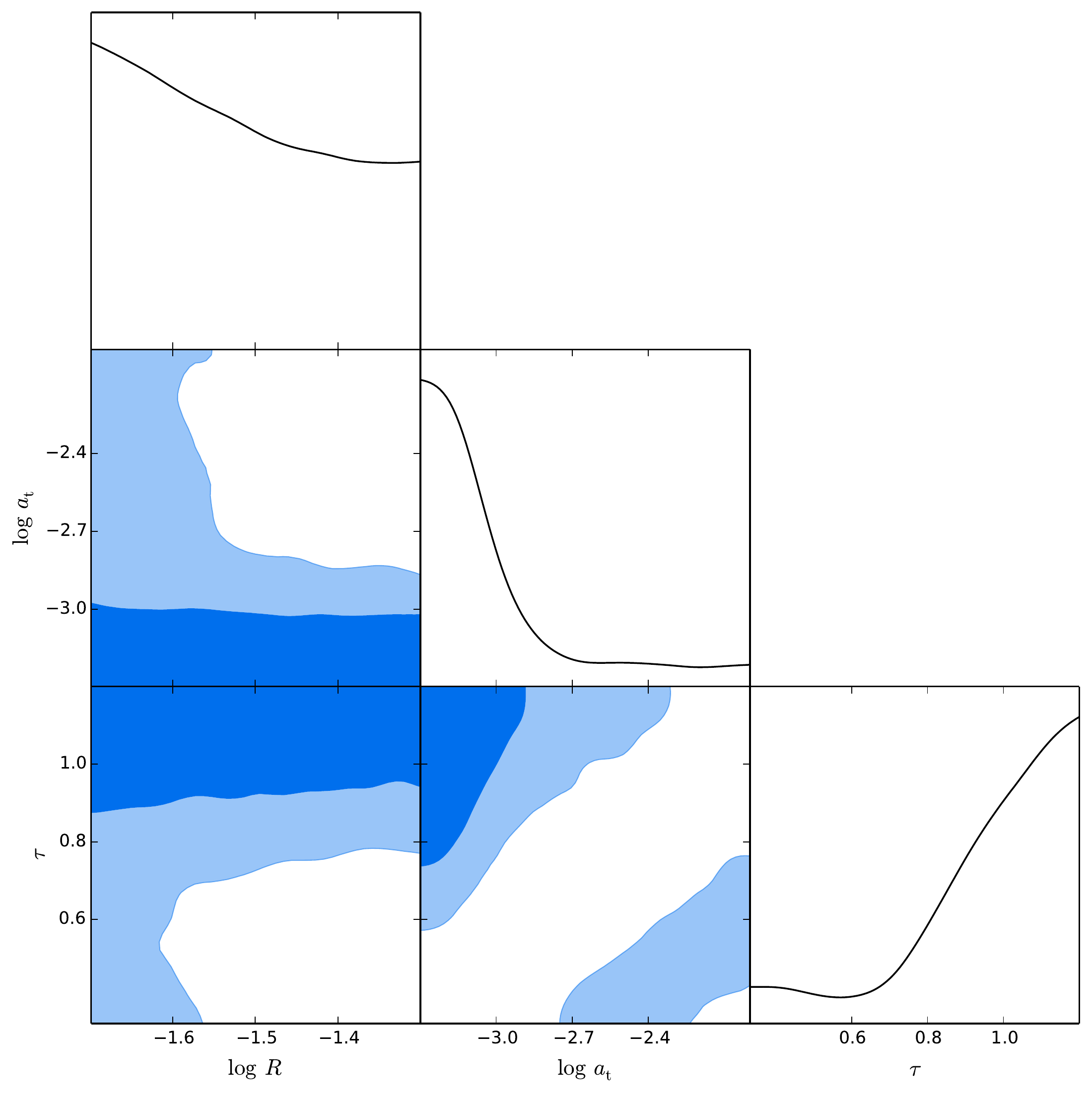} 
	\caption{As Figure~\ref{fig:plot_tri}, but restricted to $R>0.02$. 
	} 
	\label{fig:rgt02} 
\end{figure}

\begin{figure}[!thbp]
	\includegraphics[width=\columnwidth]{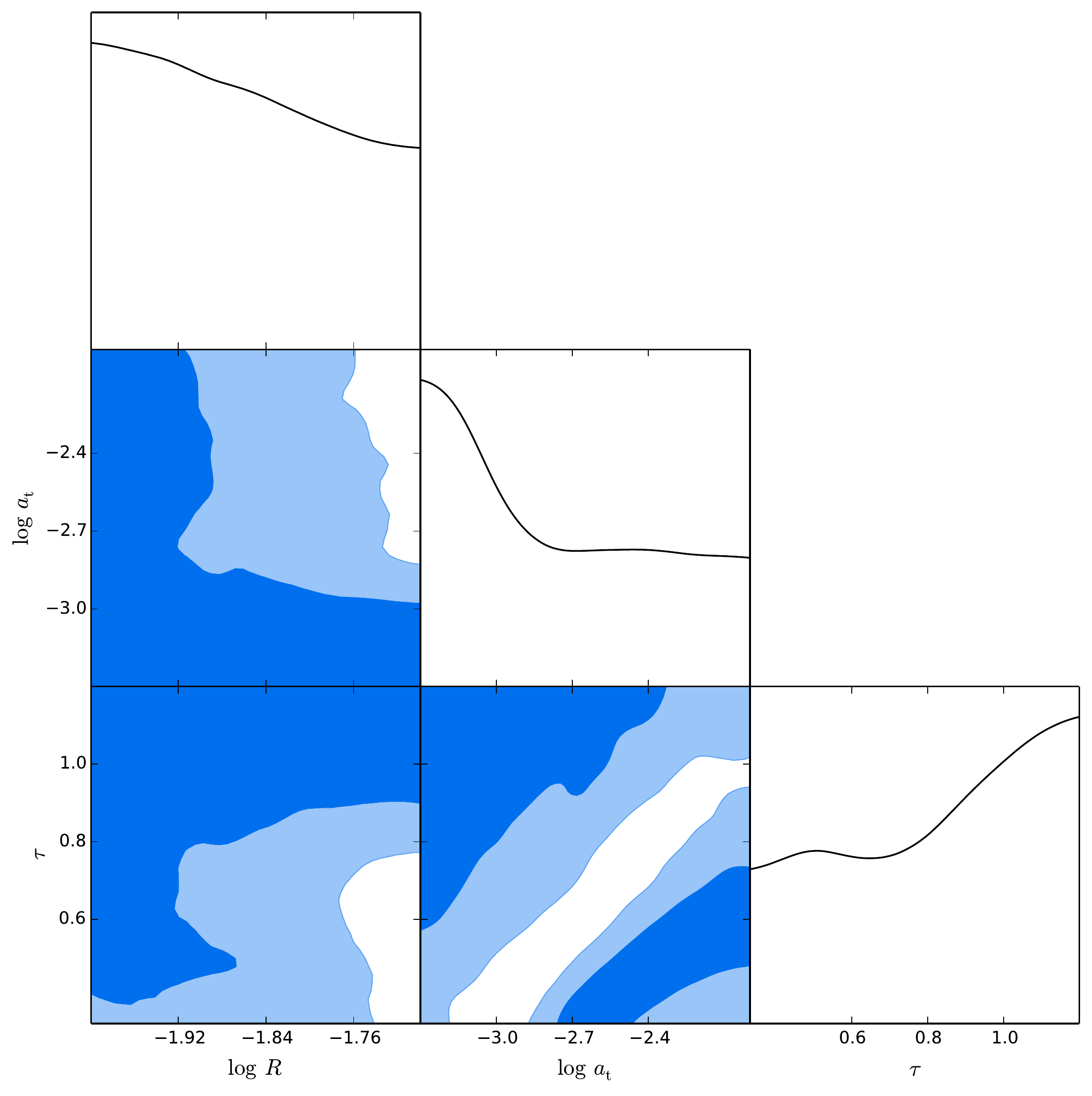} 
	\caption{As Figure~\ref{fig:plot_tri}, but restricted to $0.01<R<0.02$. 
	} 
	\label{fig:rbt0102} 
\end{figure}

This interaction between the parameters is made clearer by showing the 
confidence contours when selecting from the MCMC chains only those entries with 
various levels of $R$. Figures~\ref{fig:rgt02} and \ref{fig:rbt0102} 
illustrate the physical effects we have discussed. For example, for $R>0.02$, 
only very early $a_t$ is allowed, so the early dark energy can fade away 
appreciably by recombination (this is further helped by large $\tau$, which 
causes the fade to start earlier). For $R>0.01$, we add a region allowing 
late time transitions, but with relatively low $R$, and these can have smaller 
$\tau$ as well. Note the ``blank stripe'' in the $\log a_t$--$\tau$ plane 
starts to narrow relative to the $R>0.02$ plot as the lower $R$ means the deviation of the dark energy 
behavior from dark matter has less impact. For $R<0.01$ even more of the 
parameter space is allowed.

In summary, as a rule of thumb the data favors those regions of parameter 
space that are not too different from a $\Lambda$CDM-like cosmology. Note 
from Eqs.~(\ref{eq:neffa}) and (\ref{eq:rholss}) that a 95\% CL bound of $R\lesssim0.05$ for 
$a_t=10^{-3}$, say, implies $\neff\lesssim0.5$.

\section{Conclusions} \label{sec:concl} 

The dark sector of the Universe presents us with multiple, fundamental 
mysteries. Observations concentrate at the present epoch, where two quite 
distinct components 
appear: clustering, pressureless dark matter and highly smooth, strongly 
negative pressure dark energy. Within the visible sector of the Universe, the 
Standard Model of particle physics teaches us that apparently distinct 
entities can be unified at high energies, corresponding to early times in cosmic 
history. We have explored a phenomenological model for such a merging of 
dark matter and dark energy at early times, where barotropic aether dark energy 
has the properties of dark matter, perhaps through some unspecified direct 
interaction. At late times this behavior is gone, releasing 
the two components to evolve very differently. 

A barotropic aether model has the desired properties of naturally appearing 
like dark matter ($w=0=c_s^2$) at early times, and then a very rapid evolution
away, toward a late time attractor with $w=-1$ and $c_s^2=1$, acting like a 
cosmological constant. This would give added, rich structure to the dark 
sector and, if confirmed, a substantial clue to high energy physics. We show 
that the transition in this model cannot take place arbitrarily 
rapidly, but must take longer than a number of e-folds $\tau\ge1/3$. 

Confronting this model with current data, we find that such a model is wholly 
acceptable if the transition occurs sufficiently before recombination. For example, 
the asymptotic early ratio of dark energy to dark matter density can be 
larger than unity if the transition is at $a_t<10^{-3.84}$, or larger than 
$100$ for $a_t<10^{-4.5}$. 

Later transitions however are severely constrained by data, especially the CMB 
temperature power spectrum. 
We find that the early dark energy to dark matter density 
ratio cannot exceed 5\%, similar to other early dark energy models, for transitions 
much after recombination. Even 
later transitions have been constrained by other work, e.g.\ \cite{13052209}. 
Even so, a ratio $R=0.025$ at, say, $a_t=10^{-3}$ can contribute energy 
density interpreted in terms of an effective number of extra neutrino species 
of $\neff=0.25$. Thus early dark energy remains of interest. 

An interesting generalization is to consider a whole spectrum of barotropic 
fields, with all allowed values of sound speed $c_s^2=[0,1]$ (see 
Appendix~\ref{apx:spec}). This scenario has intriguing properties, with each component dominating 
in sequence and then fading away, similar to isotopes with different half 
lives. Because of the physical constraint $c_s^2\ge0$ the late time universe 
is left with only the $c_s^2=0$ component of dark energy, corresponding to 
the barotropic aether considered here. However, while somewhat attractive as 
a way to avoid naturally a coincidence that dark energy only dominates 
today, it does suffer from increased fine tuning (unless a way can be found to 
cancel the positive and negative contributions). 

If the dark sector does come together at high energies, we might expect the 
transition epoch not to be at eV scales ($\log a_t\approx -3$), but at GeV 
or higher scales. As this is above the primordial nucleosynthesis scale, 
observational constraints are lacking. Future work will explore whether 
inflation -- a very early dark energy period -- can constrain or benefit from 
such ``mimic'' dark energy.

\acknowledgments 

RRC thanks the IEU for hospitality during part of this work. 
This work has been supported by DOE grants DE-SC0010386 at Dartmouth, 
DE-SC-0007867 at Berkeley, and the Director, Office of Science, Office of High Energy Physics, 
of the U.S.\ Department of Energy under Contract No.\ DE-AC02-05CH11231, 
and Korea World Class University grant R32-2009-000-10130.

\appendix 

\section{A Spectrum of Barotropy} \label{apx:spec} 

Barotropic fluids can present a partial solution to the question of 
why $w\approx-1$ today, due to their rapid attraction to a de Sitter 
state.  We have also used them to influence the early expansion, through 
an unspecified interaction that makes them behave as matter in the 
early universe.  

Some attempts to solve the coincidence problem ask whether dark energy 
could have occasional influence, at several epochs early on.  This does 
not necessarily require acceleration -- which in 
any case was ruled out for the last factor $10^5$ of expansion before 
the present epoch \cite{Linder:2010wp} -- but simply to be dynamically 
relevant.  This can be done with a single scalar field with a 
sufficiently sculpted potential, e.g.\ \cite{Dodelson:2001fq,Barenboim:2005np}, or 
with a spectrum of fields \cite{Griest:2002cu,Fedrow:2013rwa,Jimenez:2012iu}.  To avoid fine 
tuning, \cite{Fedrow:2013rwa} took exponential scalar field potentials that 
trace the background energy density (keep a constant ratio, see 
\cite{Ferreira:1997au,Liddle:1998xm}), but the parameters of tracer dark energy 
necessary to give acceptable later conditions such as $w\approx-1$ then exceed 
early time observational bounds, making 
it difficult for dark energy to impact expansion at a variety of epochs. 

Here we briefly speculate about applying the idea of a spectrum of fields 
to the barotropic case.  
We emphasize that this is independent from the rest of the article, but perhaps it 
may motivate further ideas. 

Suppose we have a suite of barotropic fluids 
with a spectrum of sound speeds between 0 and 1 (recall that the sound 
speed determines the full barotropic dynamics).  Those fluids with 
$c_s^2>1/3$ may dominate in the early universe, but their energy density 
quickly redshifts away, as 
\be 
\rho_{de,j}\sim e^{-3(1+c_{s,j}^2)N}\sim a^{-3(1+c_{s,j}^2)} \ , 
\ee 
where  $N=\ln a$ is the e-folding parameter, leaving 
radiation to dominate.  We just must ensure that they fade before 
primordial nucleosynthesis. 

Those fluids with $0<c_s^2<1/3$ may affect the transition from radiation 
to matter domination, but this need not be fatal.  In fact, their energy 
density will give an effective number of neutrino species 
$N_{\rm eff}>3.046$, which may accord with the data.  They will also fade 
away as matter comes to dominate.  Again, we must make sure they fade 
before matter dominated growth is affected. 

However, since the lower limit for stable barotropic fluids is 
$c_s^2=0$, then once matter dominates the only effect from the barotropic 
dark energy arises from the constant density piece.  This effectively 
explains why there is no acceleration or dark energy influence from 
recombination until the present epoch of acceleration: barotropic dark energy 
{\it can\/} be an occasional phenomenon but once matter dominates then 
dark energy automatically appears as an approach to a de Sitter state.  
This is an attractive property of this speculative model. 

Note that we have in no way solved the fine tuning issue, since barotropic 
fluids are not tracing fields (their evolution is determined by their 
sound speed, not dynamically attracted to scale in proportion to the 
background fluid).  Indeed, we have exacerbated it since each 
fluid in the spectrum has conditions on its amplitude.  One intriguing 
possibility is that the constant density pieces of each barotropic fluid 
do not all have to be positive.  If some are positive and some negative, 
perhaps there is some way to enforce cancellation to more naturally 
end up with a small constant density. 

Thus the idea of using a spectrum of fields has some promising aspects, with 
the advantage of barotropic fluids that they quickly fade to $w\approx-1$, 
but fine tuning remains.  This is the converse of solving fine tuning 
through tracing fields (which though, since they never fade, are not 
viable).


\end{document}